\documentclass[pre,a4paper,superscriptaddress,twocolumn,showpacs,amsmath,amssymb,floatfix]{revtex4}

\usepackage{graphicx}
\usepackage{amssymb}
\usepackage{float}
\usepackage{color}

\newcommand{\bra}[1]{\langle #1 |}
\newcommand{\ket}[1]{| #1 \rangle}
\newcommand{\bracket}[2]{\langle #1| #2 \rangle}
\newcommand{\expval}[1]{\langle #1 \rangle}

\begin{document}

\title{Dynamical thermalization in Bose-Hubbard systems}

\author{Peter Schlagheck}
\affiliation{
   D{\'e}partement de Physique, University of Liege,
   4000 Li{\`e}ge, Belgium}
\author{Dima L. Shepelyansky}
%\homepage[]{http://www.quantware.ups-tlse.fr}
\affiliation{\mbox{Laboratoire de Physique Th\'eorique du CNRS, IRSAMC, 
Universit\'e de Toulouse, UPS, 31062 Toulouse, France}}

\date{today}

\begin{abstract}
We numerically study a Bose-Hubbard ring of finite size 
with disorder containing a finite number of bosons that 
are subject to an on-site two-body interaction.
Our results show that moderate interactions induce 
dynamical thermalization in this isolated system. 
In this regime the individual many-body eigenstates are well 
described by the standard thermal Bose-Einstein distribution 
for well-defined values of the temperature and the chemical 
potential which depend on the eigenstate under consideration.
We show that the dynamical thermalization conjecture
works well both at positive and negative temperatures.
The relations to quantum chaos, quantum ergodicity and to
the {\AA}berg criterion are also discussed.
\end{abstract}

\date{October 2015}


\maketitle

\section{Introduction}

A quantum system that is in contact with a thermostat is described 
by the well known quantum thermal distributions given in textbooks 
on statistical physics (see e.g.\ Ref.~\cite{landau}).
However, there had always been an interest (raised, e.g., in the
works of Bohr on the statistical description of neutron capture and 
nuclei construction \cite{bohr}) to understand the emergence of 
thermalization effects within a complex quantum system through 
the dynamical properties of the system itself, without the explicit 
introduction of a thermostat.
That is, while the full quantum system is prepared, say, within a pure
eigenstate of its Hamiltonian and is therefore not subject to thermalization, 
the one-body observables of interest in relation with its 
single-particle eigenstates may nevertheless feature the standard 
thermodynamic properties known from textbook statistical physics, as
a consequence of the presence of interactions within the system.

Of course, the emergence of such a statistical description in the 
absence of any thermostat, which we call the Dynamical Thermalization 
Conjecture (DTC) in the following, requires quantum ergodicity of the 
system eigenstates.
Research on this latter topic has been stimulated by the works 
of {\AA}berg \cite{aberg1,aberg2} as well as by Deutsch \cite{deutsch} 
and Srednicki \cite{srednicki} which generated a broad discussion of the 
Eigenstate Thermalization Hypothesis (ETH) as well as various investigations 
of different research groups (see the references and the discussion in 
Ref.~\cite{srednicki2015} and in the recent review \cite{rigol}).

It is clear that the dynamical thermalization is based on 
quantum ergodicity of eigenstates.
In the single-particle context, a mathematical proof of
quantum ergodicity was obtained by Shnirelman for eigenstates of 
one-particle chaotic billiards in the limit of large quantum numbers 
\cite{shnirelman}, where the class of billiards with chaotic 
classical dynamics had previously been established by Sinai and 
co-workers (see e.g.\ Ref.~\cite{sinai}). In such single-particle
systems it is numerically straightforward to verify that quantum ergodicity
implies the emergence of universal random matrix statistics for the 
energy levels of the system, which is known as Bohigas-Giannoni-Schmit 
conjecture established first for quantum chaos billiards \cite{bohigas}.
Thus the development of field of quantum chaos \cite{reichl,haake}
established links between quantum systems with classical dynamical chaos
and Random Matrix Theory (RMT) invented by Wigner for spectra of 
complex nuclei \cite{wigner}, and it was recognized that the emergence 
of Wigner-Dyson statistics for energy level spacings is a necessary 
condition for quantum ergodicity of eigenstates.

However, in spite of a significant progress in the field of quantum
chaos, studies in this context were mainly related to one-particle 
systems with a few degrees of freedom \cite{reichl,haake}.
Indeed, the properties of many-body quantum systems, e.g nuclei 
\cite{abohr}, were hardly accessible to computer simulations at 
the time of the 70s and 80s.
At that time the common lore within the nuclear physics community 
was that in many-body quantum systems the density of states is 
exponentially growing with the excitation energy above the Fermi level and 
hence any small interaction between fermions will very rapidly lead to the
mixing of noninteracting many-body states accompanied by the RMT 
statistics of the level spacings, by quantum ergodicity of 
states, and hence by dynamical thermalization \cite{guhr}. 
This lore persisted till the end of the 20th century even though
{\AA}berg presented in 1990 numerical and analytical arguments
according to which the onset of RMT level spacing statistics, and hence 
quantum ergodicity, takes place only when directly coupled states
are mixed by two-body interactions (which, from a fundamental point 
of view, are the only ones existing in nature) \cite{aberg1,aberg2}.
The {\AA}berg criterion for the onset of quantum chaos
in weakly interacting many-body system has been later confirmed
in more advanced studies for other quantum systems, such as
finite fermionic systems \cite{sushkov,jacquod} and quantum computers 
of interacting qubits \cite{georgeot,benenti}, as was
reviewed in Ref.~\cite{dls2001}.
In the latter context of quantum computers, examples of single 
eigenstates that are well thermalized by the presence of 
residual weak interactions between qubits and thus satisfy the 
DTC, are presented in \cite{benenti} showing that such states are also 
well described by the Fermi-Dirac thermal distribution.
In the framework of complex atoms the quantum ergodicity properties 
of eigenstates and the emergence of DTC have been discussed in 
Refs.~\cite{chirikovpla,flambaum}, but the interactions in such 
atoms are relatively strong and the DTC cannot be straightforwardly
verified in real atoms.

In the present work we investigate DTC within the physical context
of ultracold bosonic quantum gases that are confined within finite optical
lattices. Indeed, the impressive experimental progress in the handling of 
ultracold atoms and the control of their interactions has renewed and pushed
the interest in DTC and ETH and has stimulated a number of studies 
on this topic which are reviewed in Ref.~\cite{rigol} (see 
Ref.~\cite{rigol1} for a pioneering theoretical work in this context).
On the experimental side, recent investigations with cold bosonic 
atoms allow to test the validity of generalized Gibbs ensembles 
under various experimental conditions \cite{jorg}.
Even a realization of negative temperature distributions is now within
reach of cold atom experiments, as was shown in Ref.~\cite{bloch}. 
The problem of DTC and ETH is now actively investigated 
with the ultacold atoms (see e.g.\ Ref.~\cite{blochsci}).

We shall specifically consider finite Bose-Hubbard systems
with $L$ sites that contain a finite number $N$ of bosonic atoms.
Our subsystem of interest will be one (in practice arbitrarily selected)
eigenstate $\ket{k}$ of the one-body Hamiltonian describing the kinetic 
energy and the external potential within the Bose-Hubbard lattice 
(with $k \in \{1,\ldots,L\}$).
The main message that we want to convey here is that the presence
of a thermal reservoir is not necessarily required in order to achieve 
dynamical thermalization and thereby obtain the Bose-Einstein 
distribution within such a single-particle eigenstate.
We show that a moderate  (not too strong and not too weak) two-body 
interaction  $\hat{U}$, which couples the single-particle eigenstates 
with each other, can do this job as well leading to a thermal 
description of eigenstates.
In that case, the other single-particle states $\ket{k'}$ with 
$k'\neq k$ form an effective ``reservoir'' for the (sub-)``system'' 
constituted by the single state $\ket{k}$.
The temperature and the chemical potential that this reservoir provides
depend then on the specific state of the global Bose-Hubbard system 
spanned by the single-particle states $\ket{k}$, which can undergo a 
dynamical process or be prepared in one of the many-particle eigenstates 
$\ket{\Phi}$ of the full Bose-Hubbard Hamiltonian.
This latter possibility implies that a many-particle eigenstate 
$\ket{\Phi}$ of an interacting bosonic Hamiltonian can exhibit 
grand canonical thermalization features for any single-particle eigenstate 
$\ket{k}$ of its one-body (kinetic-plus-potential) part, with an effective 
temperature $T$ and an effective chemical potential $\mu$
that are specific to the state $\ket{\Phi}$.

The concept of dynamical thermalization is concretized in some more detail
in Section \ref{sec:BH} where we describe the Bose-Hubbard model under 
consideration.
Section \ref{sec:num} is devoted to presenting and discussing the
numerical results that are obtained within this Bose-Hubbard model
concerning the DTC.
Finally, possible implications of the DTC for a wider class of many-body 
systems are briefly discussed in Section \ref{sec:conc}.

\section{Dynamical thermalization within a Bose-Hubbard model}
\label{sec:BH}

Similarly as in Ref.~\cite{engl}, we consider
a one-dimensional Bose-Hubbard ring containing $L$ sites. 
The quantum many-body Hamiltonian of this system reads 
$\hat{H} = \hat{H}_0 + \hat{U}$ with
\begin{eqnarray}
  \hat{H}_0 & = & - J \sum_{l=1}^L ( \hat{a}_l^\dagger \hat{a}_{l-1} + 
  \hat{a}_{l-1}^\dagger \hat{a}_l ) 
  + \sum_{l=1}^L \epsilon_l \hat{a}_l^\dagger \hat{a}_l \,,  \label{eq:BH0} \\
  \hat{U} & = & \frac{U}{2} \sum_{l=1}^L \hat{a}_l^\dagger \hat{a}_l^\dagger 
  \hat{a}_l\hat{a}_l \,, \label{eq:BHU}
\end{eqnarray}
where $\hat{a}_l^\dagger$ and $\hat{a}_l$ respectively denote the creation
and annihilation operators associated with site $l$ and
where we formally identify $\hat{a}_0 \equiv \hat{a}_L$ and
$\hat{a}_0^\dagger \equiv \hat{a}_L^\dagger$.
The on-site energies $\epsilon_l$ ($l = 1,\ldots,L$) are fixed but randomly 
selected with uniform probability density from the interval
$-W/2 \leq \epsilon_l \leq W/2$.
Evidently, the single-particle Hilbert space of this finite system is $L$
dimensional, and the diagonalization of the single-particle Hamiltonian
corresponding to $\hat{H}_0$ therefore yields $L$ orthogonal and normalized
eigenstates $\ket{0}$, $\ket{1}$, \ldots, $\ket{L-1}$ satisfying
$\bracket{k}{k'} = \delta_{k k'}$ for all $k,k'=0,\ldots,L-1$.
The associated eigenenergies $E_k$ are supposed to be sorted such that we 
have $E_0 < E_1 < \ldots < E_{L-1}$.

Introducing the coefficients $C_{k,l}$ that handle the transformation from
the original on-site basis to the single-particle eigenbasis
through the relation
\begin{equation}
  \hat{a}_l = \sum_{k=0}^L C_{k,l}\hat{b}_k
\end{equation}
for all $l = 1, \ldots, L$, we can now represent the many-body Hamiltonian
of the Bose-Hubbard system according to $\hat{H} = \hat{H}_0 + \hat{U}$ 
with
\begin{eqnarray}
  \hat{H}_0 & = & \sum_{k=1}^L E_k \hat{b}_k^\dagger \hat{b}_k \,,
  \label{eq:H0} \\
  \hat{U} & = & \frac{U}{2} \sum_{k_1=1}^L\sum_{k_2=1}^L\sum_{k_3=1}^L\sum_{k_4=1}^L
  \sum_{l=1}^L C_{k_1,l}^* C_{k_2,l}^*C_{k_3,l}C_{k_4,l} 
  \nonumber \\ && \times
  \hat{b}_{k_1}^\dagger \hat{b}_{k_2}^\dagger \hat{b}_{k_3}\hat{b}_{k_4} \,.
  \label{eq:H}
\end{eqnarray}
In the absence of interaction, i.e.\ for $U=0$, we thereby recover the 
Hamiltonian \eqref{eq:H0} whose many-particle eigenstates are given by the
Fock states $\ket{n_0,\ldots,n_{L-1}}$ that are defined with respect to the 
single-particle basis $(\ket{0},\ldots,\ket{L-1})$.
Diagonalizing the interacting Hamiltonian $\hat{H}$ in this representation 
yields the many-body eigenstates
\begin{equation}
  \ket{\Phi_\alpha} = \sum_{n_1=0}^{L-1} \ldots \sum_{n_L=0}^{L-1} 
  \mathcal{C}_{n_0,\ldots,n_{L-1}}^{(\alpha)} \ket{n_0,\ldots,n_{L-1}}
  \label{eq:Phal}
\end{equation}
where the coefficients $\mathcal{C}_{n_0,\ldots,n_{L-1}}^{(\alpha)}$ reflect the
conservation of the total number of particles (i.e.\ we have
$\mathcal{C}_{n_0,\ldots,n_{L-1}}^{(\alpha)} = 0$ if 
$n_0 + \ldots + n_{L-1} \neq N^{(\alpha)}$ with $N^{(\alpha)}$ being the total
number of particles in the state $\ket{\Phi_\alpha}$).
We suppose that these eigenstates are ordered according to their 
associated eigenenergies $\mathcal{E}_\alpha$, i.e.\ we have 
$\mathcal{E}_0 < \mathcal{E}_1 < \mathcal{E}_2 < \ldots$.
The representation \eqref{eq:Phal} allows us now to straightforwardly 
extract the mean population of the single-particle eigenstate $\ket{k}$ 
within the many-body state $\ket{\Phi_\alpha}$ according to
\begin{eqnarray}
  \expval{\hat{n}_k}_\alpha & = & \bra{\Phi_\alpha} \hat{b}_k^\dagger \hat{b}_k 
  \ket{\Phi_\alpha} \nonumber \\
  & = & \sum_{n_1=0}^{L-1} \ldots \sum_{n_L=0}^{L-1} 
  n_k |\mathcal{C}_{n_0,\ldots,n_{L-1}}^{(\alpha)}|^2 \,. \label{eq:pops}
\end{eqnarray}

This mean population can now be compared with the prediction that would
result from a quantum statistical modeling in the spirit of the DTC. 
To this end we effectively treat each single-particle state 
$\ket{k}$ as a (sub-)system of interest and assume that the other 
states $\ket{k'}$ with $k'\neq k$ form an effective energy and particle 
reservoir to which this system is coupled by virtue of the presence of 
atom-atom interaction.
This reservoir is characterized by a given temperature $T = 1 / \beta$ 
(using temperature units in which $k_B \equiv 1$) and by a given 
chemical potential $\mu$, which both depend on the many-body eigenstate 
$\ket{\Phi_\alpha}$ under consideration, i.e.,
$\beta\equiv\beta_\alpha$ and $\mu\equiv\mu_\alpha$.

To fully establish the connection with the textbook quantum statistical
theory of noninteracting Bose gases \cite{landau}, we model the dynamics
within our system of interest by an effective one-body Hamiltonian of the 
form $\hat{H}_k^{(\rm eff)} = \tilde{E_k} \hat{b}_k^\dagger \hat{b}_k$.
While in the absence of interaction we would naturally set $\tilde{E_k} = E_k$,
the presence of a repulsive (or attractive) atom-atom interaction gives rise 
to an effective mean-field shift of this energy towards higher (or lower) 
energies.
Assuming that the atoms are more or less equidistributed among the $L$ sites
of the Bose-Hubbard ring for the many-body eigenstate under consideration
(which is generally the case in the central part of the many-body spectrum,
but may not be valid at the upper end of the spectrum in the presence of a
repulsive interaction nor at the lower end of the spectrum in the presence 
of an attractive interaction, see also the discussion in the subsequent 
section), this effective mean-field shift is approximately given by the 
addition energy $N U / L$ that would be needed in order to add an extra 
atom to the interacting system.
We therefore set 
\begin{equation}
  \tilde{E_k} = E_k + N U / L \,. \label{eq:shift}
\end{equation}

The statistical density operator of our one-state system is then written as
\begin{equation}
  \hat{\rho_k} = \frac{1}{Y_k} \exp\left[-\beta(\tilde{E}_k-\mu) 
    \hat{b}_k^\dagger \hat{b}_k \right] \,.
  \label{eq:rhok}
\end{equation}
Provided we have
\begin{equation}
  \beta (\tilde{E}_k-\mu) > 0 \,,
\end{equation}
we can express the partition function associated with the eigenstate 
$\ket{k}$ as
\begin{equation}
  Y_k = \sum_{n=0}^\infty e^{-n \beta (\tilde{E}_k-\mu)} = 
  \frac{1}{1-e^{-\beta (\tilde{E}_k-\mu)}} \,.
\end{equation}
It is then straightforward to show that the average population of the 
single-particle state $\ket{k}$ is given by the Bose-Einstein distribution
\begin{equation}
  \overline{n}_k = \mathrm{Tr}[\hat{\rho} \hat{b}_k^\dagger \hat{b}_{k}]
  = \frac{1}{e^{\beta(\tilde{E}_k - \mu)} - 1} \equiv \overline{n}_k(\beta,\mu) \,. 
  \label{eq:BEd}
\end{equation}

Applying this reasoning to all single-particle eigenstates of our Bose-Hubbard
ring gives us a means to determine the parameters $\beta$ and $\mu$ associated
with a given many-body eigenstate, provided we can trust the validity of the
DTC.
Indeed, we must have 
\begin{equation}
  \sum_{k=0}^{L-1} \overline{n}_k = N
\end{equation}
due to the conservation of the number of particles, and we can furthermore
require that the total energy of the many-body eigenstate can be evaluated as
\begin{equation}
  \mathcal{E}_\alpha = \sum_{k=0}^{L-1} \tilde{E}_k \overline{n}_k = 
  \sum_{k=0}^{L-1} E_k \overline{n}_k + N^2 U/L \,. \label{eq:Etot}
\end{equation}
These two equations can be numerically solved for $\mu$ and $\beta$.
For many-body eigenstates with a relatively low total energy 
$\mathcal{E}_\alpha$, we expect to thereby obtain a positive temperature
$T = 1/\beta > 0$ as well as a negative chemical potential satisfying
$\mu < E_k + N U / L$ for all $k = 0,\ldots,L-1$, in perfect accordance
with standard textbook quantum statistical physics \cite{landau}.
Within the upper part of the spectrum, however, we would have 
$\mu > E_k + N U / L$ for all $k = 0,\ldots,L-1$ as well as a negative 
temperature $T < 0$, which appears since the single-particle spectrum 
of the system is bounded \cite{bloch}.
We note that the concept of negative temperature is well known
in spin physics \cite{abragam},
but in our case it has a purely dynamical origin.

\section{Numerical results}
\label{sec:num}

\begin{figure}[t]
\begin{center}
\includegraphics[width=0.48\textwidth]{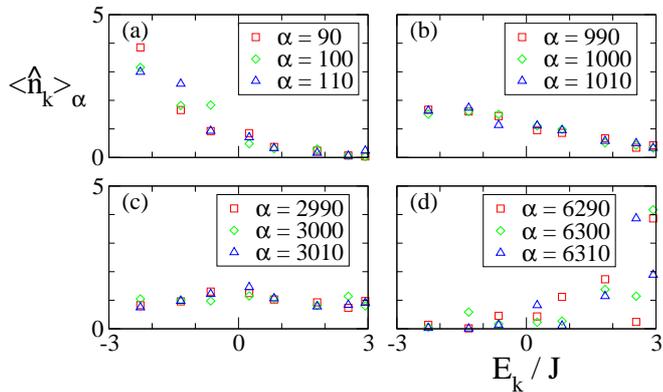}
\end{center}
\vglue -0.3cm
\caption{\label{fig1}
%\label{fig:pops_i}
Distribution of single-particle populations for various many-body 
eigenstates of the Bose-Hubbard ring with $N=8$ particles on $L=8$ sites,
with the interaction strength $U = 0.5\,J$, and with on-site energies
$\epsilon_l$ that are randomly selected within $-2 J < \epsilon_l < 2 J$.
Plotted are the populations $\expval{\hat{n}_k}_\alpha$ as a function of the 
single-particle levels $E_k$ ($k=0,\ldots,7$) for the many-body states 
$\alpha$:
(a) $\ket{\Phi_{90}}$, $\ket{\Phi_{100}}$, $\ket{\Phi_{110}}$, 
(b) $\ket{\Phi_{990}}$, $\ket{\Phi_{1000}}$, $\ket{\Phi_{1010}}$, 
(c) $\ket{\Phi_{2990}}$, $\ket{\Phi_{3000}}$, $\ket{\Phi_{3010}}$,
(d) $\ket{\Phi_{6290}}$, $\ket{\Phi_{6300}}$, $\ket{\Phi_{6310}}$.
}
\end{figure}

Fig.~\ref{fig1} displays the mean populations 
$\expval{\hat{n}_k}_\alpha$ for various many-body eigenstates that are 
obtained within a Bose-Hubbard ring of $L=8$ sites containing $N=8$ 
particles (yielding a Hilbert space that is spanned by altogether 
$\mathcal{N}=6435$ many-body eigenstates)
where we chose the parameters $U = 0.5 J$ and $W = 4 J$.
Each panel of Fig.~\ref{fig1} shows single-particle eigenstate
populations for many-body eigenstates. 
While there are weak fluctuations in the populations, the overal
probability distributions appear to remain stable with respect 
to small variations of $\alpha$, which reflects
the statistical stability of thermal distributions.

\begin{figure}[t]
\begin{center}
\includegraphics[width=0.48\textwidth]{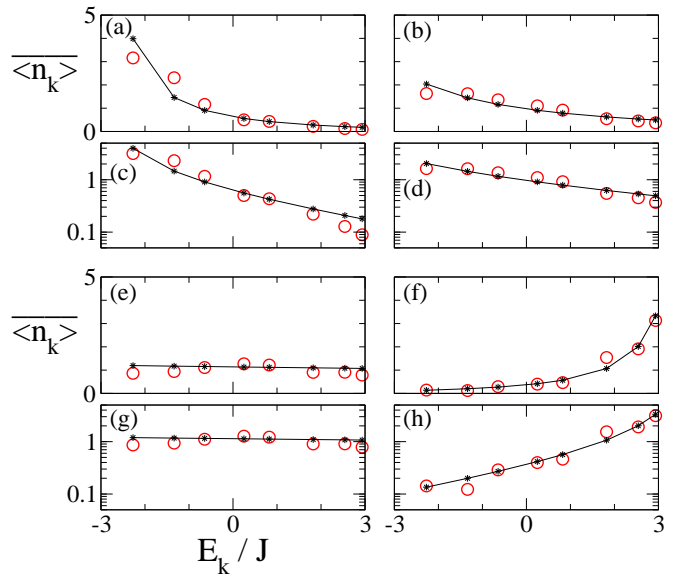}
\end{center}
\vglue -0.3cm
\caption{\label{fig2}
%\label{fig:pops_a}
Average distribution of single-particle populations for the Bose-Hubbard 
ring with $N=8$ particles on $L=8$ sites, with the interaction strength 
$U = 0.5\,J$, and with on-site energies $\epsilon_l$ that are randomly 
selected within $-2 J < \epsilon_l < 2 J$.
The red circles display, on (a,b,e,f) linear and (c,d,g,h) logarithmic scales, 
the average populations 
$\overline{\expval{n_k}} = 0.05 \sum_{\alpha = \alpha_0 - 9}^{\alpha_0 + 10}
\expval{\hat{n}_k}_\alpha$ for (a,c) $\alpha_0 = 100$, (b,d) $\alpha_0 = 1000$,
(e,g) $\alpha_0 = 3000$, and (f,h) $\alpha_0 = 6300$.
The stars connected by solid lines show the populations $\overline{n}_k$
that result from the Bose-Einstein distribution \eqref{eq:BEd} where $\beta$ 
and $\mu$ were chosen such that $\sum_k \overline{n}_k = N = 8$ and
$\sum_k E_k \overline{n}_k = \sum_k E_k \overline{\expval{n_k}}$. 
}
\end{figure}

In Fig.~\ref{fig2} we make an averaging of the populations 
$\expval{\hat{n}_k}_\alpha$ over 20 consecutive many-body eigenstates 
ranging within $\alpha_0 -9 \leq \alpha_0 \leq \alpha_0 +10$.
After such an averaging we obtain a qualitative agreement with the 
general behavior that is expected from the Bose-Einstein distribution 
\eqref{eq:BEd} shown by solid lines. 
The agreement of numerical probabilities with the DTC \eqref{eq:BEd} 
is valid for positive (panels (a,b,c,d)), infinite (e,g) and negative (f,h) 
temperatures.
In this latter case (f,h) the mean population increases with 
increasing single-particle energy, which, when being compared to 
Eq.~\eqref{eq:BEd}, would correspond to a moderately low negative
temperature $T<0$ and a chemical potential $\mu > E_{L-1}$,
while a decrease of population with increasing single-particle energy
corresponds to the more familiar case of a positive temperature 
$T>0$ and a negative chemical potential $\mu < E_0$.

\begin{figure}[t]
\begin{center}
\includegraphics[width=0.48\textwidth]{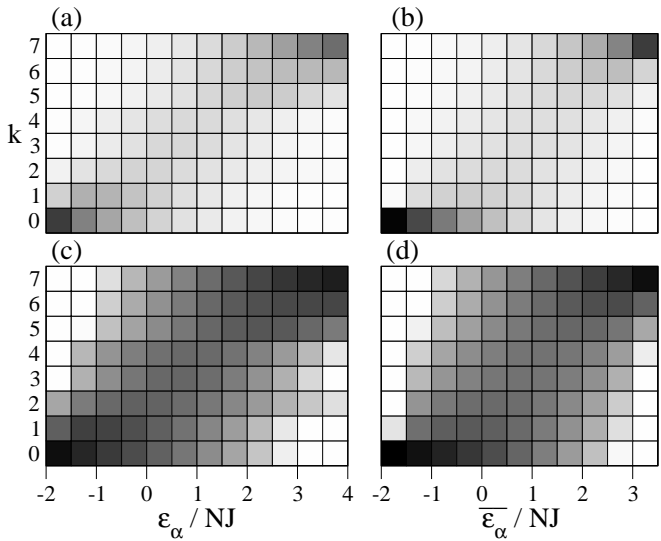}
\end{center}
\vglue -0.3cm
\caption{\label{fig3}
%\label{fig:populations}
  Left column (a,c): Average populations of the single-particle eigenstates
  $\ket{k}$ as a function of the energy per particle for the Bose-Hubbard
  ring with $N=8$ particles on $L=8$ sites, with the interaction strength 
  $U = 0.5\,J$, and with on-site energies $\epsilon_l$ that are randomly 
  selected within $-2 J < \epsilon_l < 2 J$.
  The populations are shown on a linear scale in the upper row (a,b), which 
  uniformly varies from 0 (white) to 8 (black), and on a logarithmic scale 
  in the lower row (c,d), which uniformly varies from 0.08 (white) to 8 (black).
  They are averaged over all many-body eigenstates whose energies
  per particle $\mathcal{E}_\alpha / N$ lie within the indicated intervals
  on the abscissa.
  The right column (b,d) shows the corresponding predictions provided by the
  Bose-Einstein distribution \eqref{eq:BEd} where $\beta$ 
  and $\mu$ were chosen such that $\sum_k \overline{n}_k = N = 8$ and
  $\sum_k E_k \overline{n}_k + N^2 U / L = \overline{\mathcal{E}_\alpha}$
  with $\overline{\mathcal{E}_\alpha} / N$ corresponding to the centers of
  the abscissa intervals (e.g. $\overline{\mathcal{E}_\alpha} / N = -1.75 J$
  for the leftmost sub-column of the two panels on the right-hand side).
}
\end{figure}

A more detailed comparison of the DTC with the Bose-Einstein 
distribution (\ref{eq:BEd}) is presented in Fig.~\ref{fig3}
where we show average single-particle eigenstate populations
that are obtained from all many-body eigenstates whose energies lie 
in given intervals.
The agreement between the numerically computed averages (left panels)
and the analytical predictions resulting from the Bose-Einstein distribution 
\eqref{eq:BEd} (right panels) is very good, both on a linear and on a
logarithmic scale.

\begin{figure}[t]
\begin{center}
\includegraphics[width=0.48\textwidth]{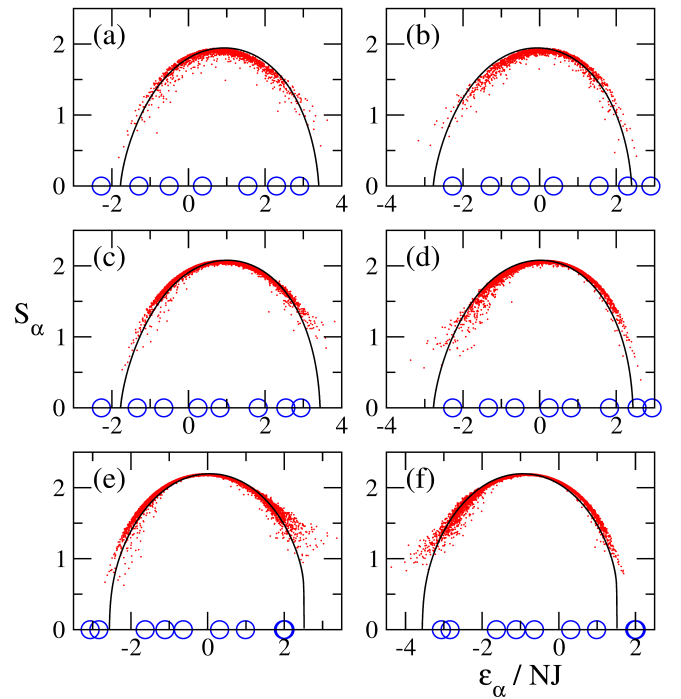}
\end{center}
\vglue -0.3cm
\caption{\label{fig4}
%\label{fig:entr}
Entropy per particle versus total energy per particle for the Bose-Hubbard
ring \eqref{eq:BH0} with on-site energies $\epsilon_l$ 
that are randomly selected within $-2 J < \epsilon_l < 2 J$.
The red dots show the entropies that are obtained from the many-body
eigenstates according to Eq.~\eqref{eq:entr} for the sizes $L=N=7$ 
(a,b), $L=N=8$ (c,d), and $L=N=9$ (e,f), and for the interaction 
strengths $U = 0.5 J$ (a,c,e) and $U = -0.5 J$ (b,d,f).
They are plotted as a function of the associated eigenenergies 
$\mathcal{E}_\alpha$ divided by the number of particles $N$.
The solid lines show the entropies that are obtained from the Bose-Einstein
distribution \eqref{eq:BEd} according to Eq.~\eqref{eq:Smu} for all
possible (positive and negative) temperatures as a function
of the corresponding energies $\overline{\epsilon}$ \eqref{eq:Emu}, 
using the single-particle eigenenergies $E_k$ of the Bose-Hubbard system,
which are displayed as (blue) circles on the abscissae.
}
\end{figure}

To assess the validity of the DTC on a more quantitative level,
we extract from the single-particle eigenstate populations 
\eqref{eq:pops} effective entropies
\begin{equation}
  S_\alpha = - \sum_{k=0}^{L-1} \frac{\expval{\hat{n}_k}_\alpha}{N}
  \ln\left( \frac{\expval{\hat{n}_k}_\alpha}{N} \right) \label{eq:entr}
\end{equation}
that characterize how many single-particle eigenstates are populated
within a given many-body eigenstate $\ket{\Phi_\alpha}$
(and that are very similar to the one-particle occupation entropies
considered in Ref.~\cite{bera} in the context of many-body localization).
Clearly, this entropy will be rather low at the lower and upper edge of
the many-body spectrum where only few single-particle states are effectively
populated (as is seen in the upper left and lower right panels of 
Fig.~\ref{fig2}, respectively), while it acquires its maximal value 
$\ln L$ in the central part of the spectrum where all single-particle 
eigenstates are equally populated on average.
Fig.~\ref{fig4} displays $S_\alpha$ as a function of the energy per 
particle $\mathcal{E}_\alpha / N$ for all many-body eigenstates of the 
Bose-Hubbard ring with $N=L=7,8,9$ and with the parameters $W = 4 J$ and 
$U = \pm 0.5 J$.
We see that there is on average a remarkable one-to-one relationship 
between the entropies $S_\alpha$ and the eigenenergies $\mathcal{E}_\alpha$ 
of the many-body eigenstates. 

The advantage of the dependence $S(E)$
is related to the fact that both variables $S$ and $E$ are
extensive variables and thus their values have smaller
fluctuations compared to probability distributions (\ref{eq:BEd}).
This feature has been noted and used for
nonlinear chains with disorder \cite{mulansky,ermann1}
and Bose-Einstein condensates, described by the Gross-Pitaivskii equation,
in chaotic two-dimensional billiards \cite{ermann2}.
It is interesting to note that in these nonlinear systems 
\cite{mulansky,ermann1,ermann2} the DTC is still valid but it 
is induced by a nonlinear mean-field interactions between 
linear states.

The entropy of probability distribution \eqref{eq:entr} can also be 
obtained in the framework of the grand canonical ensemble described 
by the Bose-Einstein distribution \eqref{eq:BEd}.
We obtain
\begin{equation}
  \overline{S}(\beta,\mu) = - \sum_k 
  \frac{\overline{n}_k(\beta,\mu)}{N(\beta,\mu)} 
  \ln\left(\frac{\overline{n}_k(\beta,\mu)}{N(\beta,\mu)}\right)
  \label{eq:Smu}
\end{equation}
with the total number of particles being given by
\begin{equation}
N(\beta,\mu) = \sum_k \overline{n}_k(\beta,\mu) \,. \label{eq:Nmu}
\end{equation}
We can then calculate $\overline{S}(\beta,\mu)$ for all possible (positive
and negative) values of $\beta$ where the chemical potential $\mu$ is
chosen such that the total population of the system according to 
Eq.~\eqref{eq:Nmu} equals the total number of particles: $N(\beta,\mu) = N$.
This population entropy can be plotted versus the effective energy per 
particle 
\begin{equation}
  \overline{\epsilon}(\beta,\mu) = \sum_k 
  \frac{\overline{n}_k(\beta,\mu)}{N(\beta,\mu)} \tilde{E}_k 
  = \sum_k \frac{\overline{n}_k(\beta,\mu)}{N(\beta,\mu)} E_k + N U/L
  \label{eq:Emu}
\end{equation}
determined in analogy with Eq.~\eqref{eq:Etot}, where we apply the 
mean-field shift \eqref{eq:shift} to account for the presence of the 
interaction.

The resulting curves are displayed by the solid lines in Fig.~\ref{fig4}.
We find a very good agreement with the population entropies $S_\alpha$
obtained from the many-particle eigenstates $\ket{\phi_\alpha}$,
for both positive and negative interaction strengths $U = \pm 0.5 J$.
Significant deviations occur near the upper bound of the spectrum for
positive interaction strengths and near the lower bound of the spectrum
for negative interaction strengths.
Indeed, many-body eigenstates in this regime are typically characterized 
by a rather strong localization of the population on a very few number
$\tilde{L} \ll L$ of Bose-Hubbard sites.
The effective mean-field shift 
$\overline{\epsilon} \mapsto \overline{\epsilon} + N U / \tilde{L}$ 
that one would have to apply in this regime is therefore much more 
important than elsewhere in the many-body spectrum.

\begin{figure}[t]
\begin{center}
\includegraphics[width=0.48\textwidth]{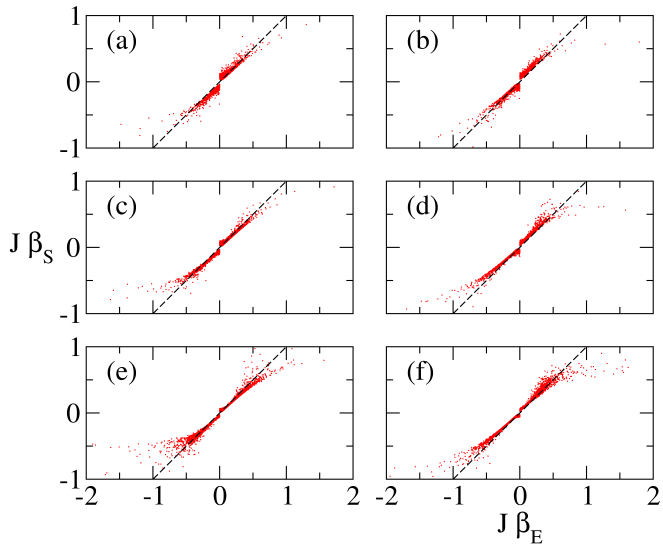}
\end{center}
\vglue -0.3cm
\caption{\label{fig5}
%\label{fig:temp}
Inverse temperatures $\beta = 1 / (k_B T)$ associated with the many-body
eigenstates of the Bose-Hubbard ring for the sizes $L=N=7$ 
(left column), $L=N=8$ (middle column), and $L=N=9$ (right column),
for the interaction strengths $U = 0.5 J$ (upper row) and 
$U = -0.5 J$ (lower row), and for on-site energies $\epsilon_l$ 
that are randomly selected within $-2 J < \epsilon_l < 2 J$,
in perfect analogy with the panels shown in Fig.~\ref{fig4}.
The horizontal axis shows the inverse temperatures $\beta_E$ that are 
obtained from intersecting the energy per particle $\mathcal{E}_\alpha/N$
of the eigenstate under consideration with the corresponding 
entropy-versus-energy curve obtained from the Bose-Einstein 
distribution (solid lines in Fig.~\ref{fig4}). 
The vertical axis shows the inverse temperatures $\beta_S$ that are 
obtained from intersecting the population entropy \eqref{eq:entr}
with the corresponding entropy-versus-energy curve.
Both possible definitions of the effective temperature of a many-body
eigenstate yield on average very similar values, as can be seen by the
fact that all data points are scattered about the diagonal (indicated
by dashed lines), with systematic deviations occurring in the regimes of 
low positive or negative temperatures to be encountered near the lower 
and upper bounds of the spectrum.
}
\end{figure}

As each point on the solid lines in Fig.~\ref{fig4} is characterized
by a well-defined temperature, we are now in a position to determine
the effective temperature $T_\alpha$ associated with a many-body eigenstate
$\ket{\phi_\alpha}$ either from its energy per particle 
$\mathcal{E}_\alpha/N$ or from its population entropy $S_\alpha$.
Both possibilities yield very similar temperatures as is seen in 
Fig.~\ref{fig5}.
Systematic deviations between these two possible definitions of the 
temperature associated with a many-body eigenstate occur in the regimes 
of low positive or negative temperatures, which are to be encountered 
near the lower and upper bounds of the spectrum.

\begin{figure}[t]
\begin{center}
\includegraphics[width=0.48\textwidth]{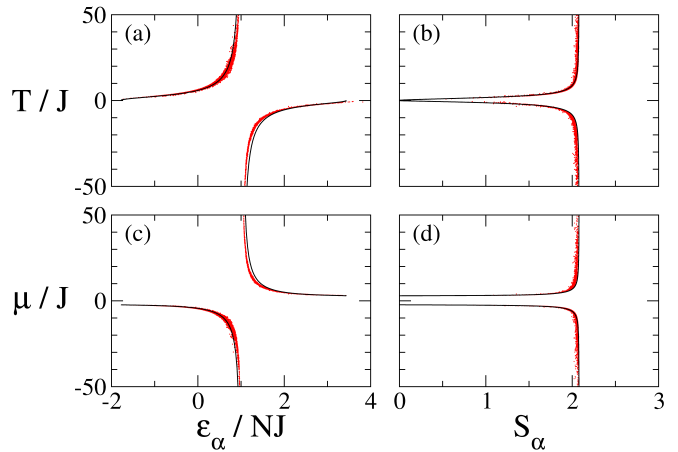}
\end{center}
\vglue -0.3cm
\caption{\label{fig6}
%\label{fig:TEMS}
Average temperatures $T$ (a,b) and chemical potentials $\mu$ (c,d)
of the many-body eigenstates as a function of their energies per particle 
$\mathcal{E}_\alpha/N$ (a,c) and their entropies $S_\alpha$ (b,d) 
for the Bose-Hubbard ring with $L = N = 8$, $U = 0.5 J$, and on-site 
energies that are randomly selected within $-2 J < \epsilon_l < 2 J$.
In the upper row (a,b), the red dots show the average temperatures 
$(\beta_E^{-1}+\beta_S^{-1})/2$ where $\beta_E$ and $\beta_S$ are taken from
Fig.~\ref{fig5}.
The red dots in the lower row (c,d) show the average chemical potentials
$(\mu_E + \mu_S)/2$ where $\mu_E$ and $\mu_S$ are obtained in a perfectly 
analogous manner as $\beta_E$ and $\beta_S$, respectively.
The solid lines display the corresponding predictions from the Bose-Einstein
distribution \eqref{eq:BEd}.
}
\end{figure}

Calculating the arithmetic average of the two possible definitions
of the temperature associated with a many-body eigenstate and plotting
this average temperature as a function of the associated energy per 
particle and entropy yields a very good agreement with the prediction
obtained from the Bose-Einstein distribution \eqref{eq:BEd}, as is seen
in Fig.~\ref{fig6}.
The same holds true for the chemical potential $\mu$ of a 
many-body eigenstate, which can also be determined either from the
corresponding energy per particle $\mathcal{E}_\alpha/N$ or the 
corresponding population entropy $S_\alpha$.
This underlines the validity of the DTC in the context of finite 
Bose-Hubbard systems.

\begin{figure}[t]
\begin{center}
\includegraphics[width=0.48\textwidth]{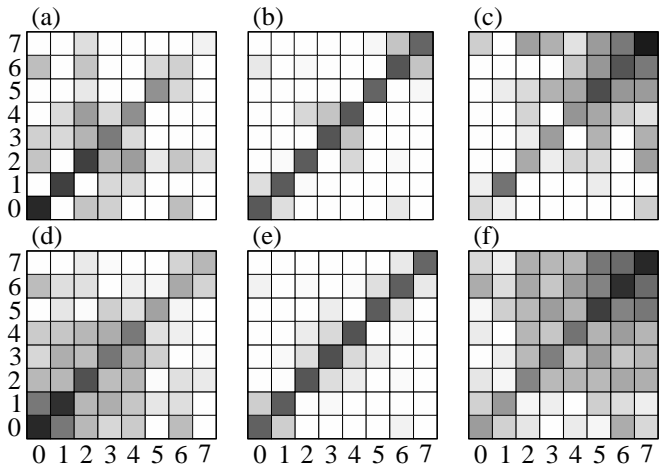}
\end{center}
\vglue -0.3cm
\caption{\label{fig7}
%\label{fig:dm}
  One-body density matrix elements associated with various many-body 
  eigenstates of the Bose-Hubbard ring with $N=8$ particles on $L=8$ 
  sites and the interaction strength $U=0.5 J$.
  The upper row (a,b,c) displays, on a logarithmic scale which uniformly 
  varies from 1 (black) to $10^{-5}$ (white), the densities 
  $p_{k,k'}^{(\alpha)}$ defined in Eq.~\eqref{eq:pkk} as a function of 
  $k$ and $k'$ (on the horizontal and vertical axis, respectively) 
  for the eigenstates (a) $\alpha = 100$, (b) $\alpha = 3000$, and 
  (c) $\alpha = 6300$.
  The lower row displays average densities
  $\overline{p}_{k,k'} = 0.01 \sum_{\alpha = \alpha_0 - 49}^{\alpha_0 + 50}
  p_{k,k'}^{(\alpha)}$ for (a) $\alpha_0 = 100$, (b) $\alpha_0 = 3000$, 
  and (c) $\alpha_0 = 6300$.
}
\end{figure}

Finally, we show in Fig.~\ref{fig7} that the reduced one-body density
matrices associated with the many-body eigenstates $\ket{\Phi_\alpha}$
are rather close to diagonal matrices.
To this end, we plot in Fig.~\ref{fig7} the densities
\begin{equation}
  p_{k,k'}^{(\alpha)} = 
  |\bra{\Phi_\alpha} \hat{b}_k^\dagger \hat{b}_{k'} \ket{\Phi_\alpha} / N |^2 \,,
  \label{eq:pkk}
\end{equation}
which correspond to the square moduli of the one-body density matrix
elements $\bra{\Phi_\alpha} \hat{b}_k^\dagger \hat{b}_{k'} \ket{\Phi_\alpha}$
normalized by the number of particles.
These densities are plotted both for individual many-body
eigenstates (upper row) and averaged over 100 consecutive eigenstates
in the many-body spectrum (lower row of Fig.~\ref{fig7}).
The logarithmic greyscale plot clearly indicates that this one-body 
density matrix is very close to a diagonal matrix.
This is precisely what one should expect to occur in the framework 
of the grand canonical ensemble defined by Eq.~\eqref{eq:rhok}

\section{Discussion}
\label{sec:conc}

In this work we demonstrated that dynamical thermalization 
takes place for interacting bosons that are contained within
a finite ring lattice exhibiting disordered on-site energies.
While the system is prepared in one of its many-body eigenstates,
the atomic populations of the single-particle eigenstates of this disordered
Bose-Hubbard system display the same thermalization features as they would
if these single-particle eigenstates were coupled to an energy and particle
reservoir according to the grand canonical ensemble.
It is therefore possible to associate to each many-body eigenstate an 
effective temperature and an effective chemical potential.
Evidently, the dynamical thermalization is caused by the presence
of the atom-atom interaction, which should not be too weak to provide an
effective mixing, nor so strong as to fully localize many-body eigenstates
on restricted spatial regions within the lattice.
In the regime of moderate interactions, individual eigenstates are well 
described by the Bose-Einstein thermal distribution satisfying DTC and ETH.

Restricting ourselves to finite size systems with up to 9 bosons,
we do not analyze in this work the conditions of validity of the DTC
in this Bose-Hubbard system. 
The determination of the conditions under which the DTC is valid 
is a much more involved task which is beyond the scope of this paper. 
In the case of long-range interactions we expect that  
the {\AA}berg createrion will work well as it was the case for
fermionic \cite{jacquod} and qubit systems \cite{georgeot,benenti}.
In the case of a finite interaction range, however, the 
noninteracting eigenstates can be localized and the validity of the DTC
can depend on the system size, on the nature and strength of 
interactions, and on the disorder strength.
The investigation of DTC and ETH in systems with a finite interaction 
range attracts now a growing interest with the possible appearance 
of phase transitions (see e.g. Ref.~\cite{alet} and references therein).
We expect that experimental tests of the DTC with cold atoms
will allow to investigate this fundamental problem in great detail.

%%%%%%%%%%%%%%%%%%%%%%%%%%%%%%%%%%%%%%%%%%%%%%%%%%%%%%%%%


\begin{thebibliography}{99}
\bibitem{landau} L.D.~Landau and E.M.~Lifshitz, 
    {\it Statistical physics}, 
    Nauka, Moskva (in Russian) (1976).
\bibitem{bohr} N.~Bohr, Nature {\bf 137}, 344 (1936).
\bibitem{aberg1} S.~{\AA}berg, Phys. Rev. Lett. {\bf 64}, 3119 (1990).
\bibitem{aberg2} S.~{\AA}berg, Prog. Part. Nucl. Phys. {\bf 28}, 11 (1992).
\bibitem{deutsch} J.M.~Deutsch, Phys. Rev. A {\bf 43}, 2046 (1991).
\bibitem{srednicki} M.~Srednicki, Phys. Rev. E {\bf 50}, 888 (1994);
         J. Phys. A {\bf 29}, L75 (1996); J. Phys. A {\bf 32}, 1163 (1999).
\bibitem{srednicki2015} K.R.~Fratus and M.~Srednicki, arXiv:1505.04296 (2015).
\bibitem{rigol} L.~D'Alessio, Y.~Kafri, A.~Polkovnikov and M.~Rigol
         arXiv:1509.06411 [cond-mat.stat-mech] (2015)
\bibitem{shnirelman} A.I.~Shnirelman, Usp. Mat. Nauk {\bf 29{6}},
     181 (1974) (in Russian).
\bibitem{sinai} I.P.~Cornfeld, S.V.~Fomin and Y.G.~Sinai, 
    {\it Ergodic theory}, 
    Springer, New York (1982).
\bibitem{bohigas} O.~Bohigas, M.J.~Giannoni and C.~Schmit, Phys. Rev. Lett.
    {\bf 52}, 1 (1984).
\bibitem{reichl} L.E.~Reichl,  
     {\it The transition to chaos: conservative classical systems and 
quantum manifestations}, Springer-Verlag, New York (2004).                
\bibitem{haake} F.~Haake, {\it Quantum signatures of chaos},
    Springer, Berlin (2010).
\bibitem{wigner} E.~Wigner, Ann. Math. {\bf 62}, 548 (1955); 
    {\bf ibid.} {\bf 65}, 203 (1957).
\bibitem{abohr} A.~Bohr and B.R.Mottelson, 
    {\it Nuclear structure}, v.1,  
    Benjamin, New York (1969).
\bibitem{guhr} T.~Guhr, A.~M\"uller-Goeling and H.A.~Weidenm\"ller,
        Phys. Rep. {\bf 299}, 189 (1999).
\bibitem{sushkov} D.~L.~Shepelyansky and O.~P.~Sushkov, 
             Europhys. Lett. {\bf 37}, 121 (1997).
\bibitem{jacquod} P.~Jacquod and D.~L.~Shepelyansky,
             Phys. Rev. Lett. {\bf 79}, 1837 (1997).
\bibitem{georgeot} B.~Georgeot and D.L.~Shepelyansky,
        Phys. Rev. E {\bf 62}, 6366 (2000).
\bibitem{benenti} G.~Benenti, G.~Casati and D.L.~Shepelyansky,
        Eur. Phys. J. D {\bf 17}, 265 (2001).
\bibitem{dls2001} D.L.~Shepelyansky, Physica Scripta {\bf T90}, 112 (2001).
\bibitem{chirikovpla} B.V.~Chirikov, Phys. Lett. A {\bf 108}, 68 (1985).
\bibitem{flambaum} V.V.~Flambaum, A.A.~Gribakina, G.F.~Gribakin
        and I.V.~Ponomarev, Physica D {\bf 131}, 205 (1999).
\bibitem{rigol1} M.~Rigol, V.~Dunjko, and M.~Olshanii, Nature {\bf 452}, 854
        (2008).
\bibitem{jorg} T.~Langen, S.~Erne, R.~Geiger, B.~Rauer, T.~Schweigler, 
         M.~Kuhnert, W.~Rohringer, I.E.~Mazets, T.Gasenzer and J.~Schmiedmayer,
         Science {\bf 348}, 207 (2015).
\bibitem{bloch} S.~Braun, J.P.~Ronzheimer, M.~Schreiber, S.S.~Hodgman,
         T.~Rom, I.~Bloch and U.~Schneider,
         Science {\bf 339}, 52 (2013).
\bibitem{blochsci} M.~Schreiber, S.S.~Hodgman, P.~Bordia, H.~L\"uschen,
                 M.H.~Fischer, R.~Vosk, E.~Altman, U.~Schneider, and I.~Bloch,
                 Science {\bf 349}, 842 (2015)
\bibitem{engl} T.~Engl, J.~Dujardin, A.~Arg\"uelles, P.~Schlagheck, 
  K.~Richter, and J.~D.~Urbina, Phys. Rev. Lett. {\bf 112}, 140403 (2014).
\bibitem{abragam} A.~Abragam, {\it The principles of nuclear magnetism},
         Oxford Univ. Press, Oxford UK (1961).
\bibitem{bera} S.~Bera, H.~Schomerus, F.~Heidrich-Meisner, and J.~H.~Bardarson,
         Phys. Rev. Lett. {\bf 115}, 046603 (2015).
\bibitem{mulansky} M.~Mulansky, K.Ahnert, A.Pikovsky and D.L.Shepelyansky,
         Phys. Rev. E {\bf 80}, 056212 (2009).
\bibitem{ermann1} L.~Ermann and D.L.~Shepelyansky, 
         New J. Phys. {\bf 15}, 12304 (2013).
\bibitem{ermann2} L.~Ermann, E.~Vergini and D.L.~Shepelyansky,
         Europhys. Lett. {\bf 111}, 50009 (2015).
\bibitem{alet} D.J.~Luitz, N.~Laflorencie and F.~Alet,
         Phys. Rev. B {\bf 91}, 081103(R) (2015). 

\end{thebibliography}
\end{document}